# Densely Connected Convolutional Networks and Signal Quality Analysis to Detect Atrial Fibrillation Using Short Single-Lead ECG Recordings

Jonathan Rubin[1*], Saman Parvaneh[1*], Asif Rahman[1], Bryan Conroy[1], Saeed Babaeizadeh[2]

[1]Philips Research North America, Cambridge, MA, USA
[2]Advanced Algorithm Research Center, Philips Healthcare, Andover, MA, USA

## Abstract

*The development of new technology such as wearables that record high-quality single channel ECG, provides an opportunity for ECG screening in a larger population, especially for atrial fibrillation screening. The main goal of this study is to develop an automatic classification algorithm for normal sinus rhythm (NSR), atrial fibrillation (AF), other rhythms (O), and noise from a single channel short ECG segment (9-60 seconds). For this purpose, signal quality index (SQI) along with dense convolutional neural networks was used. Two convolutional neural network (CNN) models (main model that accepts 15 seconds ECG and secondary model that processes 9 seconds shorter ECG) were trained using the training data set. If the recording is determined to be of low quality by SQI, it is immediately classified as noisy. Otherwise, it is transformed to a time-frequency representation and classified with the CNN as NSR, AF, O, or noise. At the final step, a feature-based post-processing algorithm classifies the rhythm as either NSR or O in case the CNN model's discrimination between the two is indeterminate. The best result achieved at the official phase of the PhysioNet/CinC challenge on the blind test set was 0.80 (F1 for NSR, AF, and O were 0.90, 0.80, and 0.70, respectively).*

## 1. Introduction

Atrial Fibrillation (AF) is the most common heart arrhythmia and its incidence in the United States alone is estimated to be 2.7-6.1 million people [1]. As such, AF screening using handheld easy-to-use devices has received a lot of attention in recent years. The goal of the 2017 PhysioNet/CinC Challenge is the development of algorithms to classify normal sinus rhythm (NSR), AF, other rhythm (O), and noisy recordings from a short single-channel ECG recording (9-60 seconds). In light of the successful utilization of deep neural networks for classification of biomedical time series (e.g., heart sounds) [2-5], a signal quality index (SQI) technique along with dense convolutional neural networks (CNN) trained with spectrogram representations were used for classification of ECG recordings.

## 2. Method and Material

A block diagram of our proposed method is shown in Figure 1. Given an ECG recording, first QRS detection takes place, followed by signal quality analysis. If the recording is judged to be of low quality (further details in Section 2.2), it is immediately classified as noisy (noise detected by SQI in Figure 1). If the recording quality is determined to be reasonable (SQI>0.5), the ECG is transformed from a one-dimensional time-series to a time-frequency representation and consecutively evaluated using one of two CNNs, depending on the signal recording length. The first model accepts as input 15-second ECG segments. However, if the input recording is shorter than 15 seconds, a secondary model that processes 9-second ECG segments is used.

Both models employ the Densely Connected Convolutional Network (DenseNet) architecture [6]. Compared to a standard CNN architecture, each layer within a DenseNet architecture concatenates all preceding layer feature-maps as input. Figure 2 illustrates this concept, where arrows indicate reused feature-maps from previous layers in a five-layer dense block.

Each DenseNet model accepts as input a spectrogram segment computed by consecutive Fourier transforms (details could be found in section 2.4). The original DenseNet architecture was modified to ensure batch normalization [7] could be performed row-wise (i.e. normalizing over frequency bins per batch). This modification resulted in networks that outperformed standard channel-wide batch normalization.

If the input ECG was labelled as NSR or O by the DenseNet model, an additional check was performed by a post-processing unit (further details in Section 2.5).

---

*[*] Authors contributed equally*

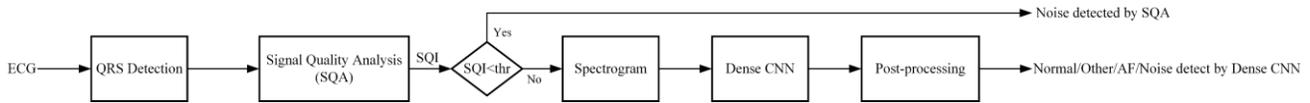

Figure 1. Block diagram of the proposed algorithm.

## 2.1. Data Splitting and Augmentation

The training set for the challenge included 8,528 single-channel ECG recordings (NSR: 5050, AF: 738, other rhythm: 2456, and noisy: 284). Details about the challenge dataset can be found in [8]. A 5-fold stratified split was applied to the 8,528 ECG recordings made available by the challenge organizers. Stratified splitting was used to maintain class prevalence between the data splits. Recordings from four of the splits were used to construct a training/validation set (6821 ECG recordings) made up of the QRS aligned spectrogram segments. The training set included 80% of the above recordings. The other 20% were used as a validation set during model training. The remaining stratified split, consisting of 1,707 ECG recordings, was kept as an in-house test-set for assessing algorithm performance, independent from the blind challenge test dataset.

A further 6,312 30-second ECG segments representing atrial fibrillation were collected from various sources (including ambulatory recordings from Holter monitors) and used to augment the training and validation sets. Baseline wander was removed from each AF segment and was up-sampled from 250 to 300 samples-per-second in order to match the sampling rate of the challenge dataset.

## 2.2. QRS Detection and Signal Quality Assessment

After removing baseline wander using a moving average filter, QRS complexes were detected using gqrs algorithm, publicly available in WFDB toolbox [9]. After aligning by the detected QRS peaks, average template matching correlation coefficient [10] with the threshold of 0.5 was used as SQI to identify noisy data. This measure had the highest area under the receiver operating characteristic (ROC) curve for discriminating between artefacts and arrhythmic ECG [11].

## 2.3. Spectrogram

For each recording, a spectrogram was constructed using an FFT applied on a moving window with the length of 75 samples and overlap of 50%. Segments with the length of 15 and 9 seconds were extracted from the spectrogram beginning at each of the detected QRS peaks.

## 2.4. Dense Convolutional Neural Networks

If the quality of ECG recording was reasonable (SQI> 0.5) by the SQI module, rhythm classification took place using a dense convolutional neural network. Recordings processed by CNNs were classified as NSR, AF, O, as well as noisy. Recall that, at first, an attempt is made to use a CNN model that processes 15-second segments. However, if the input recording length is not long enough, a secondary model that processes 9-second segments is used.

### 2.4.1. Main and Secondary Models

**Main model**: The 15-second model is made up of 3 dense blocks consisting total of 40 layers. Each layer involves applying a convolutional filter, followed by ReLU activation and row-wise batch normalization. A growth rate of 6 feature-maps was used for each layer. Model input dimensions were a single channel of 20 frequency bins by 375 time segments. The first 20 frequency bins from the computed spectrogram captures a frequency range of up to approximately 50 Hz. In total, the model consisted of 262,344 trainable parameters.

**Secondary Model**: The architecture used for the secondary model was similar to the main model, however, a smaller growth rate of 4 feature-maps was used per layer. Model input dimensions were a single channel of 20 frequency bins by 225 time segments, height and width. The lower width resulting from the shorter 9-second segment size. In total, the secondary model consisted of 119,458 trainable parameters.

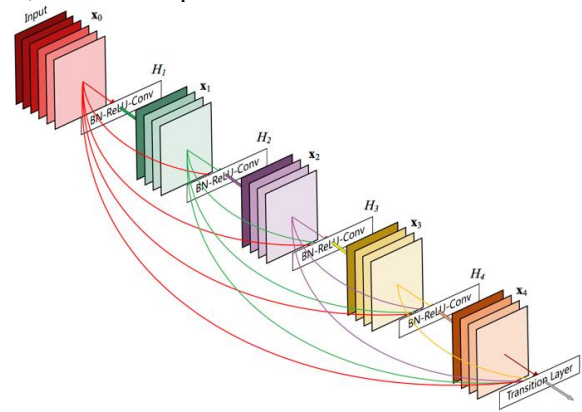

Figure 2. Five layers of a DenseNet block with a growth rate of 4 feature-maps per layer (source [6]).

### 2.4.2. Model Training

Both the main and secondary models were trained as four class classification models using standard softmax cross-entropy loss. Models were typically trained for no more than 15 epochs. Once a model was sufficiently trained, in-house testing was performed on the left-out stratified split, as previously described. Models that achieved desirable performance were further trained before submission to the challenge server. In particular, the full five splits of challenge data were used to train a final model, where 95% of the data was used for training and the remaining 5% for validation. Final model training did not occur from scratch, but rather weights from the previously learned model were used to pre-initialize the dense CNN for continued training using the updated, full dataset.

### 2.5. Post Processing

If the ECG is labelled as NSR or O by the CNN and the probability of being NSR and O are close to each other (absolute difference between probability of NSR and O < 0.4), a feature-based post-processing step is performed to cast the final decision. For NSR/O post-processing, an AdaBoost-abstain classifier [12] was trained using the NSR and O recordings in the in-house training set. Its performance was tested on the in-house test set. A total of 437 features were extracted from five different categories to train the model:
- Signal quality (2 features): average template matching correlation coefficient [10] and bSQI [13] based on the output of gqrs and Pan-Tompkins [14] QRS detection algorithms.
- Frequency content (10 features): median power across nine frequency bands (1-15, 15-30, 30-45, 45-60, 60-75, 75-90, 90-150, 5-14, and 5-50Hz) as well as ratio of power in 5-14Hz band to power in 5-50Hz. The power spectrum of the ECG record was estimated using discrete-time Fourier transform.
- Beat to beat interval (11 features): number, minimum, maximum, and median of RR intervals, SDNN, RMSSD, average heart rate, and different heart rate asymmetry measures (PI, GI, SI)
- ECG-based reconstructed phase space (401 features): normalized ECG reconstructed phase space (RPS) was created with dimension 2 and delay equal to 4 samples [15]. Then, the RPS was divided into small square areas (grid of 20×20). Normalized number of points in each square was considered a feature. In addition, spatial filling index was calculated [16].
- Poincare section from ECG (13 features): using RPS reconstructed from ECG, as described above, 13 different features from Poincare section with unity line were extracted. More details about the method and features can be found in [15, 17].

### 2.6. Algorithm Evaluation

Performance of the algorithm was evaluated using an average of three F1 values for classification of NSR, AF, and O ($F1_n$, $F1_a$, and $F1_o$, respectively). In-house test set was used for algorithm evaluation independent from the blind challenge test dataset. Also, performance was tested on a random subset of blind hidden test during official phase and final score was created using the whole blind test set.

### 3. Results

Area under the ROC curve for AdaBoost–abstain classifier in NSR/O post-processing step was 0.86 on the in-house test set. Only 58 features were selected by the classifier, the top 10 were from beat to beat interval (n=5), ECG-based reconstructed phase space (n=2), and Poincare section from ECG (n=3).

The best result achieved of the proposed algorithm at the official phase of the challenge on the in-house test set was 0.82 (F1 for NSR, AF, and other rhythm were 0.91, 0.80, and 0.76, respectively). Final result on the whole challenge blind test dataset was 0.80 (F1 for NSR, AF, and other rhythm were 0.90, 0.80, and 0.70, respectively).

Table 1. Algorithm performance on in-house test set and whole blind test set. $F1_n$, $F1_a$, and $F1_o$ are F1 values for classification of NSR, AF, and O.

| Dataset | $F1_n$ | $F1_a$ | $F1_o$ | F1 |
|---|---|---|---|---|
| In-house test set | 0.91 | 0.80 | 0.76 | 0.82 |
| Whole blind test set | 0.90 | 0.80 | 0.70 | 0.80 |

### 4. Discussion

This work led to the development and evaluation of several model types, not all of which are fully described in this paper. Here we discuss some of the findings of this effort, as well as alternative approaches investigated during the CinC challenge.
- One of the findings of this work was that for CNNs that process spectrograms as input, row-wise batch normalization (i.e. normalizing over frequency bins per batch) is preferable to a typical channel-wide application of batch normalization. This modification to our CNN models consistently resulted in considerable performance gains.
- Significant experimentation was performed using so called *wide and deep networks* [18], where activations from later convolutional layers (deep features) are combined with variables that capture information using domain knowledge (wide features). The wide features that were considered included well-known HRV measurements (e.g. SDNN, RMSSD, pNN50), entropy measure (e.g. SampleEn) and morphological features (e.g. P-wave duration, PR interval, QT-interval). However, the addition of wide features typically resulted in approximately a 2% drop in overall performance. It is possible that additional wide features, not presently included within our experimentation, would result in performance improvement.

- Lastly, the model presented here achieved its current performance using only time-frequency inputs encoded as spectrograms. A further model type was tested that accepted time-frequency inputs (as described), as well as a parallel CNN architecture that accepted the raw ECG waveform as input to automatically capture morphological information. These two parallel models were combined to make a final classification. This dual network that captured frequency and morphology information showed promise on our in-house test set results – producing $F_1$ scores that outperformed all other network architectures that were evaluated. However, these networks resulted in computational requirements that were beyond the restrictions imposed by the challenge server, hence we were not able to assess their overall performance on the hidden challenge test dataset.

## 5. Conclusion

In this article, a SQI technique was combined with dense convolutional neural networks following by a post-processing feature-based classifier to find the best method for distinguishing atrial fibrillation from normal sinus rhythm, other rhythms, and noise. The promising performance of the algorithm makes us hopeful that with further enhancement this technique may be suitable for practical clinical use.

## References

[1] C. T. January, L. S. Wann, J. S. Alpert, H. Calkins, J. C. Cleveland, J. E. Cigarroa*, et al.*, "2014 AHA/ACC/HRS guideline for the management of patients with atrial fibrillation," *Circulation,* p. CIR. 0000000000000041, 2014.
[2] C. Potes, S. Parvaneh, A. Rahman, and B. Conroy, "Classifier ensemble for detection of abnormal heart sounds," 2016.
[3] J. Rubin, R. Abreu, A. Ganguli, S. Nelaturi, I. Matei, and K. Sricharan, "Recognizing Abnormal Heart Sounds Using Deep Learning," *arXiv preprint arXiv:1707.04642,* 2017.
[4] J. Rubin, R. Abreu, A. Ganguli, S. Nelaturi, I. Matei, and K. Sricharan, "Classifying heart sound recordings using deep convolutional neural networks and mel-frequency cepstral coefficients," in *Computing in Cardiology Conference (CinC), 2016*, 2016, pp. 813-816.
[5] C. Potes, S. Parvaneh, A. Rahman, and B. Conroy, "Ensemble of feature-based and deep learning-based classifiers for detection of abnormal heart sounds," in *Computing in Cardiology Conference (CinC), 2016*, 2016, pp. 621-624.
[6] G. Huang, Z. Liu, K. Q. Weinberger, and L. van der Maaten, "Densely connected convolutional networks," *arXiv preprint arXiv:1608.06993,* 2016.
[7] S. Ioffe and C. Szegedy, "Batch normalization: Accelerating deep network training by reducing internal covariate shift," in *International Conference on Machine Learning*, 2015, pp. 448-456.
[8] G. Clifford, C. Liu, B. Moody, I. Silva, Q. Li, A. Johnson*, et al.*, "AF Classification from a Short Single Lead ECG Recording: the PhysioNet Computing in Cardiology Challenge 2017," presented at the Computing in Cardiology Rennes-France, 2017.
[9] I. Silva and G. B. Moody, "An open-source toolbox for analysing and processing physionet databases in matlab and octave," *Journal of open research software,* vol. 2, 2014.
[10] C. Orphanidou, T. Bonnici, P. Charlton, D. Clifton, D. Vallance, and L. Tarassenko, "Signal-quality indices for the electrocardiogram and photoplethysmogram: derivation and applications to wireless monitoring," *IEEE journal of biomedical and health informatics,* vol. 19, pp. 832-838, 2015.
[11] C. Daluwatte, L. Johannesen, L. Galeotti, J. Vicente, D. Strauss, and C. Scully, "Assessing ECG signal quality indices to discriminate ECGs with artefacts from pathologically different arrhythmic ECGs," *Physiological measurement,* vol. 37, p. 1370, 2016.
[12] B. Conroy, L. Eshelman, C. Potes, and M. Xu-Wilson, "A dynamic ensemble approach to robust classification in the presence of missing data," *Machine Learning,* vol. 102, pp. 443-463, 2016.
[13] J. Behar, J. Oster, Q. Li, and G. D. Clifford, "ECG signal quality during arrhythmia and its application to false alarm reduction," *IEEE transactions on biomedical engineering,* vol. 60, pp. 1660-1666, 2013.
[14] J. Pan and W. J. Tompkins, "A real-time QRS detection algorithm," *IEEE transactions on biomedical engineering,* pp. 230-236, 1985.
[15] S. Parvaneh, M. R. Hashemi Golpayegani, M. Firoozabadi, and M. Haghjoo, "Predicting the spontaneous termination of atrial fibrillation based on Poincare section in the electrocardiogram phase space," *Proceedings of the Institution of Mechanical Engineers, Part H: Journal of Engineering in Medicine,* vol. 226, pp. 3-20, 2012.
[16] O. Faust, R. Acharya, S. Krishnan, and L. C. Min, "Analysis of cardiac signals using spatial filling index and time-frequency domain," *BioMedical Engineering OnLine,* vol. 3, p. 30, 2004.
[17] S. Parvaneh, M. R. H. Golpaygani, M. Firoozabadi, and M. Haghjoo, "Analysis of Ecg In Phase Space for the Prediction of Spontaneous Atrial Fibrillation Termination," *Journal of electrocardiology,* vol. 49, pp. 936-937, 2016.
[18] H.-T. Cheng, L. Koc, J. Harmsen, T. Shaked, T. Chandra, H. Aradhye*, et al.*, "Wide & deep learning for recommender systems," in *Proceedings of the 1st Workshop on Deep Learning for Recommender Systems*, 2016, pp. 7-10.

Address for correspondence.
Jonathan Rubin / Saman Parvaneh
2 Canal Park, 3rd floor, Cambridge, MA 02141.
jonathan.rubin@philips.com / saman.parvaneh@philips.com